\documentclass[aps,prl,twocolumn,superscriptaddress,twoside,floatfix,amsmath,showpacs,letterpaper]{revtex4}

\usepackage{graphicx}

\newcommand{\lnsco}{La$_{1.6-x}$Nd$_{0.4}$Sr$_x$CuO$_4$}
\newcommand{\lnscoxy}{La$_{2-x-y}$Nd$_{y}$Sr$_x$CuO$_4$}

\newcommand{\pcco}{Pr$_{2-x}$Ce$_{x}$CuO$_{4+\delta}$}
\newcommand{\ybco}{YBa$_{2}$Cu$_{3}$O$_{y}$}

\newcommand{\cecusix}{CeCu$_{6-x}$Au$_x$}

\begin{document}

\title{Thermopower across the pseudogap critical point of \lnsco :\\
Evidence for a quantum critical point in a hole-doped high-$T_c$ superconductor}

\author{R. Daou}
\affiliation{D\'{e}partement de physique and RQMP, Universit\'{e} de
  Sherbrooke, Sherbrooke, Canada}
  
\author{Olivier Cyr-Choini\`{e}re}
\affiliation{D\'{e}partement de physique and RQMP, Universit\'{e} de
  Sherbrooke, Sherbrooke, Canada}
  
\author{Francis Lalibert\'{e}}
\affiliation{D\'{e}partement de physique and RQMP, Universit\'{e} de
  Sherbrooke, Sherbrooke, Canada}
  
\author{David LeBoeuf}
\affiliation{D\'{e}partement de physique and RQMP, Universit\'{e} de
  Sherbrooke, Sherbrooke, Canada}
  
\author{Nicolas Doiron-Leyraud}
\affiliation{D\'{e}partement de physique and RQMP, Universit\'{e} de
  Sherbrooke, Sherbrooke, Canada}
  
\author{J.-Q. Yan}
\affiliation{Texas Materials Institute, University of Texas at Austin,
  Austin, Texas 78712, USA}

\author{J.-S. Zhou}
\affiliation{Texas Materials Institute, University of Texas at Austin,
  Austin, Texas 78712, USA}

\author{J.B. Goodenough}
\affiliation{Texas Materials Institute, University of Texas at Austin,
  Austin, Texas 78712, USA}
 
\author{Louis Taillefer}
\altaffiliation{E-mail: louis.taillefer@physique.usherbrooke.ca}
\affiliation{D\'{e}partement de physique and RQMP, Universit\'{e} de
  Sherbrooke, Sherbrooke, Canada}
\affiliation{Canadian Institute for Advanced Research, Toronto, Canada}

\date{\today}

\begin{abstract}

The thermopower $S$ of the high-$T_c$ superconductor \lnsco{} was measured
as a function of temperature $T$ near its pseudogap critical point, the 
critical hole doping $p^\star$ where the pseudogap temperature $T^\star$ 
goes to zero. 
Just above $p^\star$, $S/T$ varies as $\ln(1/T)$ over a decade of 
temperature. Below $p^\star$, $S/T$ undergoes a large increase below 
$T^\star$. 
As with the temperature dependence of the resistivity, which is 
linear just above $p^\star$ and undergoes a large upturn below 
$T^\star$, these are typical signatures of a quantum phase transition.
This suggests that $p^\star$ is a quantum critical point below which
some order sets in, causing a reconstruction of the Fermi surface, 
whose fluctuations are presumably responsible for the linear-$T$ resistivity 
and logarithmic thermopower.
We discuss the possibility that this order is the ``stripe'' order known to
exist in this material.

\end{abstract}



\maketitle


The nature of the pseudogap phase in high-$T_c$ superconductors has yet to
be elucidated. Quantum oscillation studies \cite{doiron07} have revealed that
the large hole-like Fermi surface characteristic of highly overdoped
cuprates \cite{hussey03} is modified in the pseudogap phase, where it contains
small electron-like pockets \cite{leboeuf07}. A fundamental question is: what
causes this change in Fermi surface? Is it the onset of some order?
If so, what symmetry is broken?

\begin{figure}
\centering
\includegraphics[width=0.48\textwidth]{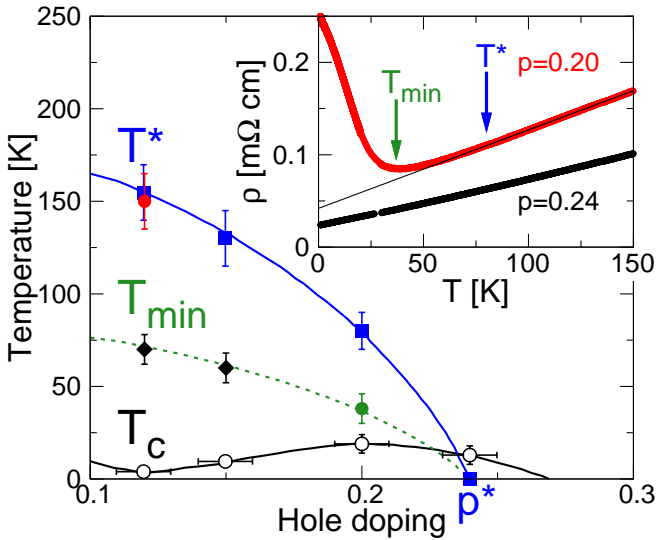}
\caption{(Color online) Phase diagram of Nd-LSCO. The pseudogap temperature $T^\star$
  (blue squares and solid line) is defined as the temperature below which the 
  normal-state resistivity deviates from its linear-$T$ behavior at 
  high temperature (see inset and \cite{daou08}). 
  This is in agreement with the (extrapolated) opening
  of the pseudogap seen by photoemission at $p=0.12$ (red circle)~\cite{chang08}.
  Since the linear-$T$ dependence extends down to $T \to 0$ at $p=0.24$ (see inset), 
  $T^\star = 0$ at that doping. Although we define the pseudogap critical point 
  $p^\star$ to be at $p = 0.24$, it could be slightly lower.
  The superconducting transition temperature $T_c$ (open black circles) is also plotted, 
  showing that the $T^\star$ line must end inside the superconducting phase. 
  Data for $p=0.12$ and $p=0.15$ are from~\cite{ichikawa00}; data for $p=0.20$ and 
  $p=0.24$ are from~\cite{daou08}.
  Also shown is the upturn temperature 
  $T_{\rm min}$ (closed green circle and dashed line) at which the resistivity reaches its minimum value (see inset).
  The onset of charge order deduced from X-ray diffraction~\cite{x-ray98,niemoller99} 
  (black diamonds) coincides with $T_{\rm min}$ (see Refs.~\cite{daou08,ichikawa00}).
  Inset: 
  normal-state resistivity of the two Nd-LSCO crystals used in this study, measured
  in a magnetic field strong enough to entirely suppress superconductivity (from~\cite{daou08}).}
\label{fig:phasediag}
\end{figure}

\begin{figure}
\centering
\includegraphics[width=0.48\textwidth]{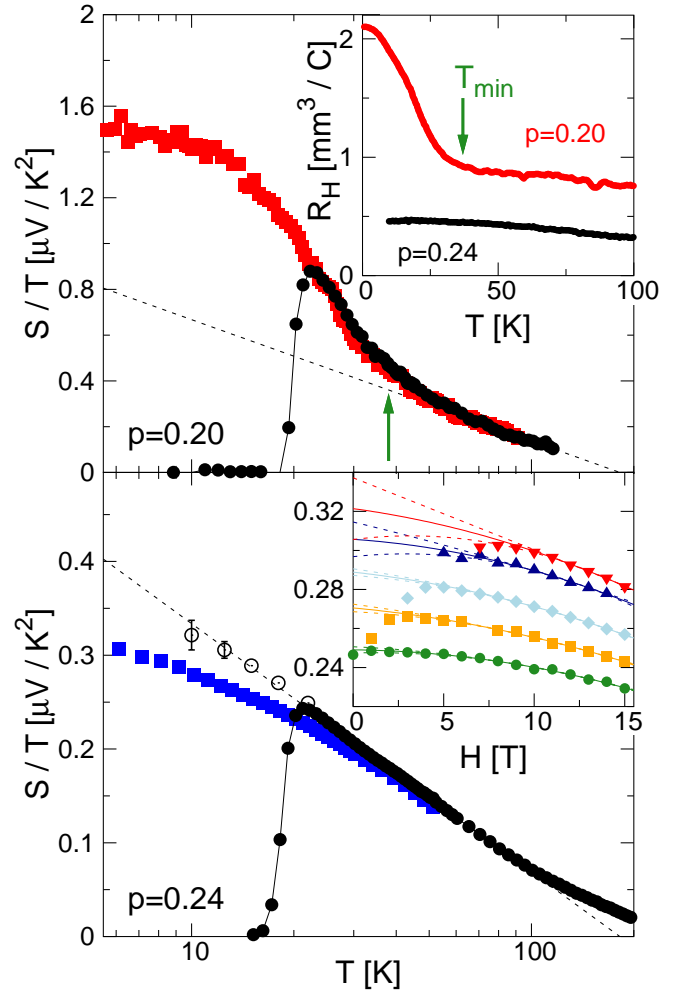}
\caption{(Color online) Thermopower $S(T)$ of Nd-LSCO, plotted as $S/T$ vs log$T$, with (squares)
  and without (full circles) a magnetic field of 15 T applied along the $c$-axis.
  Top panel: 
  sample with $p=0.20$, measured with the DC technique (see text). 
  The magnetic field has no discernible effect other than to suppress
  superconductivity. 
  The dashed line is a linear fit to the data above 50\,K.
  The arrow marks $T_{\rm min}$ (see inset of Fig.~\ref{fig:phasediag}).
  Inset: 
  Hall coefficient as a function of temperature, for both Nd-LSCO
  samples~\cite{daou08}.
  At $p = 0.20$, the upturn in $R_H(T)$ at low temperature is seen
  to coincide with $T_{\rm min}$.
  Bottom panel: 
  sample with $p=0.24$, measured with the AC technique (see text).
  The magnetic field is seen to cause a reduction of $S$ for $T < 40$~K.
  To correct for this and extend the zero-field behavior to $T < T_c$, 
  we extrapolate finite field data to zero field (see inset). The resulting 
  extrapolated values are plotted as open circles.
  %
  Inset:
  $S/T$ for Nd-LSCO with $p=0.24$ as a function of magnetic 
  field, at five fixed temperatures: 
  10, 12.5, 15, 18 and 22 K (top to bottom).  
  Second order polynomial fits to the field dependence are extrapolated 
  back to zero field. Best (solid lines) and worst (dashed lines) fits are shown, 
  indicative of the uncertainty in the width and position of the superconducting downturn.
  The corresponding error bars are shown in the main panel (open circles).}
  
\label{fig1}
\end{figure}

Recent measurements of the
normal-state resistivity $\rho(T)$ and Hall coefficient $R_H(T)$ in
the hole-doped cuprate \lnsco ~(Nd-LSCO) have shown that the change in Fermi surface
in this material occurs at the critical doping $p^\star$ where the pseudogap 
temperature $T^\star$ goes to zero \cite{daou08}, as illustrated in
Fig.~\ref{fig:phasediag}. 
Recent photoemission measurements on Nd-LSCO at $p = 0.12$ have
confirmed that the pseudogap in Nd-LSCO has the same features as
in other cuprates \cite{chang08}, with an onset temperature consistent with
$T^\star$ determined from transport (see Fig.~\ref{fig:phasediag}).
At a hole doping $p = 0.20$, 
$\rho(T)$ shows a pronounced upturn below $T^\star$ (see inset to
Fig.~\ref{fig:phasediag}). The simultaneous upturn in $R_H(T)$ is strong
evidence that these upturns are caused by a change in the Fermi
surface \cite{daou08}. At a
slightly higher doping, $p = 0.24$, $R_H(T)$ remains flat at
low temperature (see inset in top panel of Fig.~\ref{fig1}), with the value
expected of a large hole-like Fermi surface containing $1 + p$ holes
~\cite{daou08}. In that large-Fermi-surface state, $\rho(T)$ is linear in
temperature down to the lowest temperatures \cite{daou08}
(see inset to Fig.~\ref{fig:phasediag}).

In this Letter, we investigate the thermopower $S(T)$ of Nd-LSCO. 
In general, the thermopower is a complex quantity that involves the energy 
dependence of the conductivity ~\cite{behnia04,miyake05}. 
However, in the limit of dominant impurity scattering, it has been shown 
theoretically that $S/T \propto (C_e/T)(1/ne)$, where $C_e$ is the electronic 
specific heat, $n$ is the density of charge carriers and $e$ is the charge of the 
electron \cite{miyake05}.
Empirically, it has been pointed out that $S/T \approx (C_e/T)(1/ne)$ in the limit of
$T \to 0$ for a wide range of strongly correlated electron systems \cite{behnia04}.
Therefore, at low temperature the thermopower approximately
represents the electronic heat capacity per charge carrier.
(Note that it would be difficult to measure $C_e(T)$ accurately in
Nd-LSCO given that it is less than 1~\% of the total specific heat $C(T)$ 
above 4~K, and the low-temperature behavior is 
masked by a large Schottky anomaly \cite{takeda01,sutjahja03}.)
We find that the three regimes of behavior seen in the resistivity as upturn 
for $p < p^\star$, linear for $p = p^\star$ and quadratic for $p > p^\star$, show up in $S/T$ 
respectively as upturn, logarithmic divergence and nearly flat. 
This is strongly reminiscent of the electron behavior in metals
near a quantum phase transition \cite{lohneysen07}, suggesting
that the pseudogap phase is characterized by some order,
which vanishes at a quantum critical point located inside the region of 
superconductivity in the phase diagram. 
Evidence points to so-called ``stripe'' order, as the anomalies in transport 
coincide with the onset of spin / charge modulations.
 
The two samples of Nd-LSCO used in this study are the same as used and 
described in Ref.~\cite{daou08}. They have a doping of $p=0.20$ and $p=0.24$, with respective 
$T_c$ values of 20\,K and 17\,K. 
The thermopower was measured using a one-heater, two-thermometer 
DC technique, with Cernox thermometers. The applied temperature
gradient was always less than 7\,\% of the average 
sample temperature. 
The thermopower of the resistive leads in the measurement circuit 
(PtW or phosphor-bronze) was calibrated against
optimally-doped \ybco{} ($T_c = 93$~K) for $T < 90$\,K
and 6N-pure Pb for $T > 90$\,K \cite{roberts77}. 
The $p = 0.24$ sample was also measured using a low-frequency 
two-heater, two-thermometer AC technique \cite{chen01}, with a
sinusoidal excitation of frequency 5--100\,mHz and amplitude 0.1\,K. 
The signal-to-noise ratio in the AC measurement was 10 times better
than in the DC case. There was excellent agreement in the data obtained 
with both techniques.


In Fig.~\ref{fig1}, we plot $S/T$ vs log$T$ for Nd-LSCO at $p = 0.20$ and
$p=0.24$. 
The data taken on our $p = 0.20$ crystal ($x = 0.20$ and $y = 0.4$)
is in excellent quantitative agreement with previous measurements on \lnsco{} at the
same values of $x$ and $y$, over the entire temperature range in zero 
magnetic field~\cite{nakamura92}.
To our knowledge, no in-field thermopower data has been reported on Nd-LSCO, 
nor is there any published data on Nd-LSCO for $x > 0.20$.
Our data on Nd-LSCO at $p = 0.24$ is in good quantitative agreement with published 
data on polycristalline LSCO at $p = 0.25$ (only reported in zero field) \cite{uher1987,elizarova2000}.


There is no consensus on the mechanism that governs the thermopower in cuprates.
While phonon drag has been invoked to
explain the temperature dependence in Bi-2212~\cite{trodahl95}, it is not satisfactory
for the case of Nd-LSCO and YBCO where neither electron-phonon nor
mass-enhancement mechanisms are adequate~\cite{zhou95,tallon95}. 
Here we propose an electronic origin for both the temperature and
doping dependence of $S$, at least below 100 K.
This is strongly supported by the similarity found in resistivity and
Hall effect.

At a doping $p=0.24$, close to $p^\star$, $S/T$ in zero magnetic field
shows a perfect $\ln(T_0/T)$ dependence from 100~K down to $T_c$.
%
%
Application of a magnetic field $H \parallel c = 15$~T to push $T_c$ 
down is seen to slightly suppress $S/T$ below this $\ln(T_0/T)$ dependence for
$T < 40$\,K. By extrapolating the field dependence of $S/T$ to $H = 0$, as
shown in the inset to Fig.~\ref{fig1}, we can track the zero-field $S/T$ at
temperatures below $T_c(H=0)$. This shows that the $\ln(1/T)$ regime persists at
least down to 10\,K, within the uncertainty of this extrapolation, {\it i.e.}
over a full decade of temperature.


\begin{figure}
\centering
\includegraphics[width=0.48\textwidth]{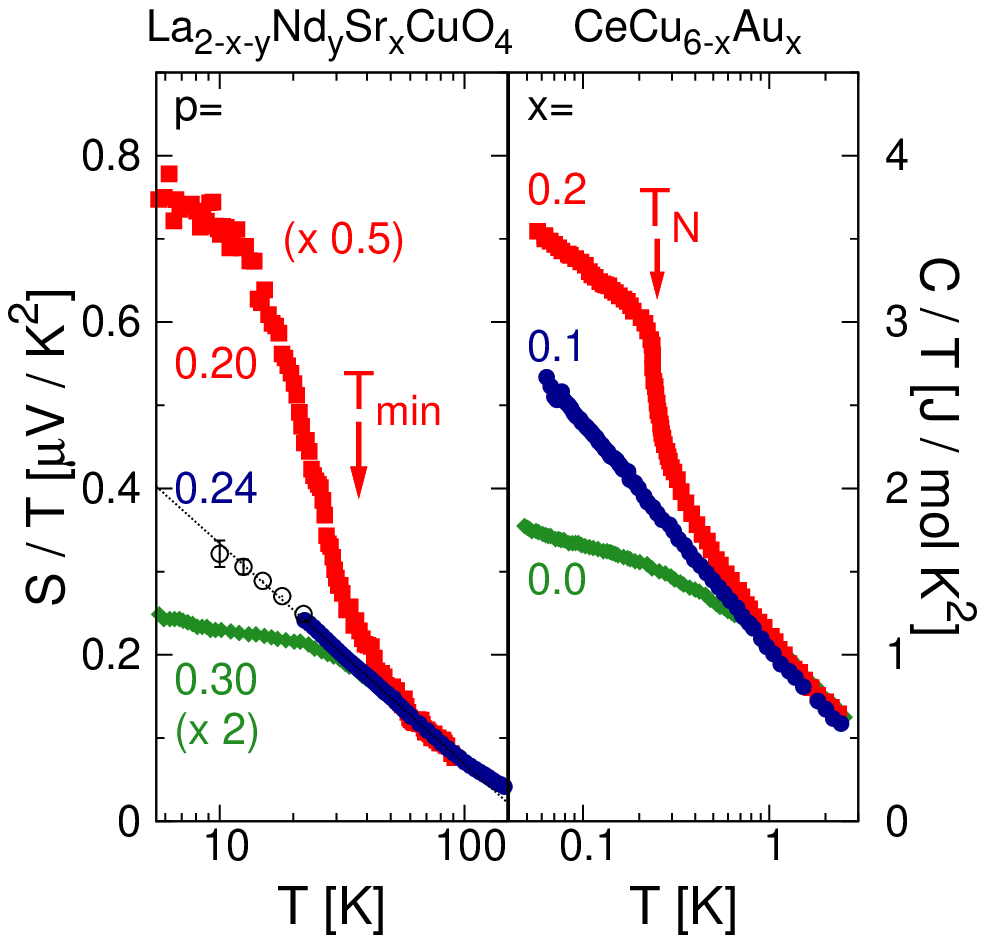}
\caption{(Color online) 
  Left panel: 
  Thermopower of \lnscoxy{} with $p=0.20$ ($\times$~0.5) and
  $p=0.24$~($y=0.4$, this work), compared to that of LSCO with $p=0.30$ 
  ($\times$~2; $y=0$, from \cite{kamran}), plotted as $S/T$~vs log$T$.
  %
  Right panel: 
  Specific heat of the heavy-fermion metal \cecusix{} at $x=0.0$, 
  $0.1$ and $0.3$, plotted as $C_e/T$~vs log$T$, 
  showing the evolution across the quantum critical point 
  at $x = x_c = 0.1$ where the
  ordering temperature $T_N$ goes to zero (from \cite{lohneysen94}). 
  %
  }
\label{fig3}
\end{figure}

This $\ln(T_0/T)$ dependence of $S/T$ is strongly reminiscent of the 
$\ln(T_0/T)$ dependence observed in $C_e/T$ at the quantum critical point 
of various heavy-fermion metals \cite{lohneysen07}. 
In Fig.~\ref{fig3}, we compare $S/T$ in Nd-LSCO with $C_e/T$ in the heavy-fermion
metal \cecusix ~\cite{lohneysen94}, each at three concentrations:
below, at and above their respective critical points, $p^\star$ and $x_c$.
By substituting Au in CeCu$_6$, antiferromagnetic order is made to
appear beyond a critical concentration $x = x_c = 0.1$, with an ordering temperature 
$T_{\rm N}$ that rises linearly with $x$ ~\cite{lohneysen94,lohneysen07}. 
In the absence of data on Nd-LSCO at $p > p^\star$, we compare with data on LSCO ($y=0$)
at $x = p = 0.30$~\cite{kamran}. Given that both
materials exhibit virtually identical
resistivity and thermopower above $T^\star$ \cite{nakamura92}, it is reasonable to 
assume they also do above $p^\star$.

\begin{figure}
\centering
\includegraphics[width=0.48\textwidth]{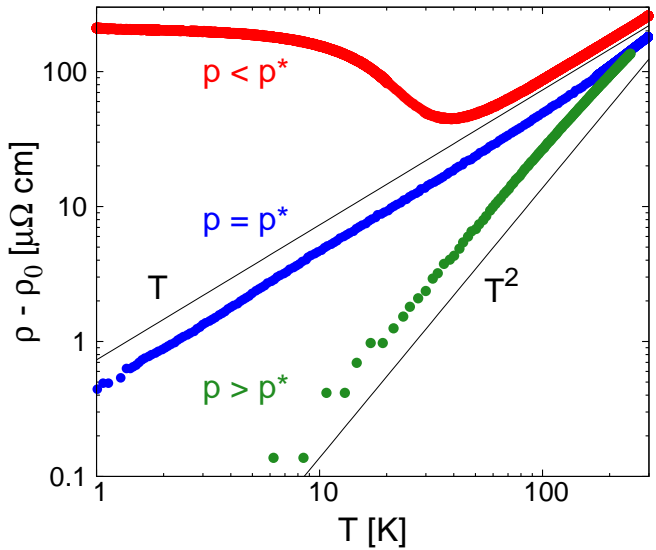}
\caption{(Color online) Temperature dependent part of the resistivity,
  $\rho(T) - \rho_0$, vs log$T$ for \lnsco ~with $p=0.20$ 
  ($p < p^\star$) and $p=0.24$~($p = p^\star$), from \cite{daou08},
  compared to that of LSCO with $p=0.30$ ($p > p^\star$), 
  from \cite{nakamae03}.
  $\rho_0$ is the value to which $\rho(T)$ extrapolates at $T = 0$; 
  for Nd-LSCO at $p = 0.20$, the extrapolation is based only
  on data above $T^\star = 80$~K.}

  %
  %
  
\label{fig2}
\end{figure}

The similarity is remarkable, both materials
displaying the three distinctive regimes of quantum criticality:
relatively flat in the Fermi-liquid state, logarithmically divergent
at the critical point, and a jump in the ordered state. 
The characteristic temperature scale $T_0$ in the $\ln(T_0/T)$ dependence of
either $S/T$ or $C_e/T$ is of course vastly different in the two materials, 
by roughly two orders of magnitude, as are the ordering and pseudogap temperatures,
$T_N$ and $T^\star$. 
This qualitative similarity reinforces the case for a quantum phase transition in Nd-LSCO 
at $p^\star$, previously made on the basis of resistivity~\cite{daou08}, 
whose three regimes are displayed in Fig.~\ref{fig2}:
quadratic in the Fermi-liquid state,
linear at the critical point, and an upturn below that point.


There is also a strong similarity with the electron doped cuprate \pcco~(PCCO),
where the case for a quantum critical point is well established \cite{dagan04}.
In the $T \to 0$ limit, both $R_H$ and $S/T$ in PCCO show an abrupt change
as the doping $x$ drops below the critical doping $x_c$, signalling the 
change in Fermi surface from a large hole cylinder to a combination of small 
electron and hole pockets \cite{li07,lin05}.
The two coefficients track each other, 
as equivalent measures of the effective carrier density~\cite{li07}.
At $x =x_c$, $\rho(T)$ is again linear in temperature at low $T$ \cite{fournier98}.
%
%
%
These typical signatures of a quantum critical point have been attributed to the loss of 
antiferromagnetic order near $x_c$~\cite{motoyama07}, and the quantum fluctuations
thereof.
 
In a model of charge carriers on a three-dimensional Fermi surface
scattered by two-dimensional antiferromagnetic spin fluctuations, 
transport properties near the magnetic quantum critical point are found to be dominated
by ``hot spots'', points on the
Fermi surface connected by the ordering wavevector.
In this case, calculations show that
$\rho(T) \propto T$, $C_e/T \propto \ln(T_0/T)$
and
$S/T \propto \ln(T_0/T)$, where $k_B T_0$ is an energy scale 
on the order of the bandwidth \cite{paul01}.
This naturally accounts for the different temperature scales observed in
Nd-LSCO and \cecusix{} where $T_0 \simeq 170$~K in the former and 4\,K in
the latter, since the Fermi velocity is about $10^5$~m/s in cuprates and
$10^3$~m/s in heavy-fermion metals.

The strong empirical similarity with both heavy-fermion metals and
electron-doped cuprates makes a compelling case for a quantum critical 
point at $p^\star$ in Nd-LSCO.
The nature of the order below $p^\star$ seems to involve both spin and charge 
degrees of freedom.
On the one hand, superlattice Bragg peaks observed in Nd-LSCO by neutron diffraction
show that a static (or slow) spin modulation at low temperature persists all the
way up to $p \approx p^\star$~\cite{ichikawa00}.
On the other hand, the upturn in $\rho(T)$ at $T_{\rm min}$ coincides with the
onset of charge order~\cite{daou08}, which occurs at a temperature
somewhat above the onset of spin modulation~\cite{ichikawa00}. 
In other words, the pseudogap phase below $T^\star$ (and $p^\star$) appears to be a phase with
``stripe'' order, perhaps short-range or fluctuating above $T_{\rm min}$.

The impact of stripe order on the Fermi surface of a hole-doped
cuprate has been calculated~\cite{millis07}. 
The large hole-like cylinder is found to reconstruct in a way that depends
on the strength of the spin and charge  
potentials. Calculations of the associated Hall coefficient predict a
rise in $R_H$ with the onset of charge order~\cite{lin08}, as observed
experimentally in Nd-LSCO when going from 
$p=0.24$ to $p=0.20$~\cite{daou08}. 
Spin order can cause a drop in $R_H$, which can even become negative~\cite{lin08},
as a result of an electron pocket being present in the reconstructed
Fermi surface~\cite{millis07}. Such a drop is indeed seen in Nd-LSCO at
lower doping, in the vicinity of $p = 1/8$, where $R_H(T \to 0) \approx
0$~\cite{nakamura92}.   
The fact that a large drop in $R_H(T)$ also occurs in YBCO near $p = 1/8$~\cite{leboeuf07}, 
starting at a very similar temperature, points to a common underlying 
cause of Fermi-surface reconstruction.


In conclusion, the combination of resistivity, Hall coefficient and 
thermopower in Nd-LSCO makes a compelling case that the pseudogap phase
in this high-$T_c$ superconductor ends at a quantum critical point 
located inside the superconducting dome at $p \approx 0.24$.
All three transport coefficients undergo a simultaneous rise below a
temperature $T_{\rm min}$ which coincides with the onset of charge order
seen by other probes. This strongly suggests that the Fermi surface
is reconstructed by ``stripe'' order. 
Given that a linear-$T$ resistivity is a universal
property of cuprates near optimal doping, it is likely that a common 
mechanism is at play, associated with such a quantum critical point, 
in analogy with heavy-fermion metals.
%

\acknowledgments

We thank K. Behnia and N.E. Hussey for allowing us to show their
unpublished thermopower data on LSCO at $x=0.30$~\cite{kamran}.
%
We also thank A. Chubukov, P. Coleman, Y.B. Kim, S.A. Kivelson, G. Kotliar, 
K. Haule, G.G. Lonzarich, A.J. Millis, M.R. Norman, C. Proust, T.M. Rice, S. Sachdev, 
T. Senthil, H. Takagi and A.-M.S. Tremblay for helpful discussions, and J. Corbin 
for his assistance with the experiments. 
LT acknowledges the support of a Canada Research Chair, NSERC, FQRNT and CIfAR. 
JSZ and JBG were supported by an NSF grant.


\begin{thebibliography}{99}

\bibitem{doiron07} N. Doiron-Leyraud {\it et al.}, Nature {\bf 447}, 565 (2007).

\bibitem{hussey03} N.E. Hussey {\it et al.}, Nature {\bf 425}, 814 (2003).

\bibitem{leboeuf07} D. LeBoeuf {\it et al.}, Nature {\bf 450}, 533 (2007).

\bibitem{daou08} R. Daou {\it et al.}, Nature Phys. (DOI:10.1038/nphys1109).

\bibitem{chang08} J. Chang {\it et al.}, New J. Phys. {\bf 10}, 103016 (2008).

\bibitem{ichikawa00} N. Ichikawa {\it et al.}, Phys. Rev. Lett. {\bf 85}, 1738 (2000).

\bibitem{x-ray98} M.v. Zimmermann {\it et al.}, Eur. Phys. Lett. {\bf 41}, 629 (1998).

\bibitem{niemoller99} T. Niem\"{o}ller {\it et al.}, Eur. Phys. J. B {\bf 12}, 509 (1999).

\bibitem{behnia04} K. Behnia {\it et al.}, J. Phys. Cond. Mat. {\bf 16}, 5187 (2004).

\bibitem{miyake05} K. Miyake and H. Kohno, J. Phys. Soc. Jpn {\bf 74}, 254 (2005).

\bibitem{takeda01} J. Takeda {\it et al}., Phys. Chem. Sol. {\bf 62}, 181 (2001).

\bibitem{sutjahja03} I.M. Sutjahja {\it et al.}, Physica C {\bf 392-396}, 207 (2003).

\bibitem{lohneysen07} H.v. L\"{o}hneysen {\it et al.}, Rev. Mod. Phys. {\bf 79}, 1015 (2007).


\bibitem{roberts77} R.B. Roberts, Phil. Mag. {\bf 36}, 91 (1977).

\bibitem{chen01} F. Chen {\it et al.}, Rev. Sci. Instr. {\bf 72}, 4201 (2001).

\bibitem{nakamura92} Y. Nakamura and S. Uchida, Phys. Rev. B {\bf 46}, 5841 (1992).

\bibitem{uher1987} C. Uher {\it et al.}, Phys. Rev. B {\bf 36}, 5676 (1987).

\bibitem{elizarova2000} M.V. Elizarova {\it et al.}, Phys. Rev. B {\bf 62}, 5989 (2000).

\bibitem{trodahl95} H.J. Trodahl, Phys. Rev. B {\bf 51}, 6175 (1995).

\bibitem{zhou95} J.-S. Zhou and J.B. Goodenough, Phys. Rev. B {\bf 51}, 3104 (1995).

\bibitem{tallon95} J.L. Tallon {\it et al.}, Phys. Rev. Lett. {\bf 75}, 4114 (1995).

\bibitem{lohneysen94} H.v. L\"{o}hneysen {\it et al.},  Phys. Rev. Lett. {\bf 72}, 3262 (1994).

\bibitem{kamran} H. Jin, A. Narduzzo, M. Nohara, H. Takagi, N.E. Hussey
  and K. Behnia, to be published. 

\bibitem{nakamae03} S. Nakamae {\it et al.},  Phys. Rev. B {\bf 68}, 100502(R) (2003).

\bibitem{dagan04} Y. Dagan {\it et al.}, Phys. Rev. Lett. {\bf 92}, 167001 (2004).

\bibitem{li07} P. Li, K. Behnia and R.L. Greene, Phys. Rev. B {\bf 75}, 020506(R) (2007).

\bibitem{lin05} J. Lin and A.J. Millis, Phys. Rev. B {\bf 72}, 214506 (2005).

\bibitem{fournier98} P. Fournier {\it et al.}, Phys. Rev. Lett. {\bf 81}, 4720 (1998).

\bibitem{motoyama07} E.M. Motoyama {\it et al.}, Nature {\bf 445}, 186 (2007).

\bibitem{paul01} I. Paul and G. Kotliar, Phys. Rev. B {\bf 64}, 184414 (2001).


\bibitem{millis07} A.J. Millis and M.R. Norman, Phys. Rev. B {\bf 76}, 220503(R) (2007).

\bibitem{lin08} J. Lin and A.J. Millis, Phys. Rev. B {\bf 78}, 115108 (2008).

\end{thebibliography}
\end{document}